\documentclass[conference]{IEEEtran}
\usepackage{cite}
\usepackage{amsmath,amssymb,amsfonts}
\usepackage{algorithmic}
\usepackage{graphicx}
\usepackage{textcomp}
\usepackage{xcolor}
\usepackage{amsmath,amssymb,amsfonts}
\usepackage{textcomp}
\usepackage{graphicx}
\usepackage{times}
\usepackage{microtype}
\usepackage[UKenglish]{babel}
\usepackage{graphicx,dblfloatfix}
\usepackage{blindtext}
\usepackage{mathtools}
\usepackage{multirow}
\usepackage{tabularx}
\usepackage{subcaption}
\usepackage{hhline}
\usepackage{arydshln}
\usepackage{listings}
\usepackage{adjustbox}
\usepackage{color,soul}
\usepackage[margin=1in]{geometry}
\begin{document}

\title{Exploring Parallelism in FPGA-Based Accelerators for Machine Learning Applications \\

\thanks{Funding info hidden for double-blind submission.}
}

\author{
    \IEEEauthorblockN{Sed Centeno\IEEEauthorrefmark{1}, Christopher Sprague\IEEEauthorrefmark{1}, Arnab A Purkayastha\IEEEauthorrefmark{1}, Ray Simar\IEEEauthorrefmark{2}, Neeraj Magotra\IEEEauthorrefmark{1}}
    \IEEEauthorblockA{\IEEEauthorrefmark{1}Western New England University, Springfield, MA, USA\\
    sed.centeno, christopher.sprague, arnab.purkayastha, neeraj.magotra@wne.edu}
    \IEEEauthorblockA{\IEEEauthorrefmark{2} Rice University, Houston, TX, USA\\
    ray.simar@rice.edu}
}

\maketitle

\begingroup
\renewcommand\thefootnote{}\footnote{\textbf{This work has been accepted for publication in the proceedings of IEEE MWSCAS 2025. This is the author’s preprint version. The final published version will be available on IEEE Xplore.}}
\endgroup


\begin{abstract} 


Speculative backpropagation has emerged as a promising technique to accelerate the training of neural networks by overlapping the forward and backward passes. Leveraging speculative weight updates when error gradients fall within a specific threshold reduces training time without substantially compromising accuracy. In this work, we implement speculative backpropagation on the MNIST dataset using OpenMP as the parallel programming platform. OpenMP's multi-threading capabilities enable simultaneous execution of forward and speculative backpropagation steps, significantly improving training speed. The application is planned for synthesis on a state-of-the-art FPGA to demonstrate its potential for hardware acceleration.

Our CPU-based experimental results demonstrate that speculative backpropagation achieves a maximum speedup of 24\% in execution time when using a threshold of 0.25, and accuracy remaining within 3-4\% of the baseline across various epochs. Additionally, when comparing individual step execution time, speculative backpropagation yields a maximum speedup of 35\% over the baseline, demonstrating the effectiveness of overlapping forward and backward passes.

\end{abstract}
 
\vspace*{0.5 cm}

\section{Introduction} 
\label{sec:Introduction}


The rapid proliferation of machine learning (ML) applications has brought forth significant computational challenges, particularly in training large and complex models. The computational complexity and extensive training time of deep learning algorithms necessitate the development of faster and more efficient hardware accelerators. Traditionally, Graphics Processing Units (GPUs) have been the dominant hardware choice for ML workloads due to their high throughput and parallelism. However, Field Programmable Gate Arrays (FPGAs) have emerged as a compelling alternative due to their ability to offer customizable parallel processing and low power consumption.

Several studies have shown that FPGAs can benefit from the replication of Single Instruction, Multiple Thread (SIMT) execution model, commonly associated with GPUs \cite{socctaxonomy, dnn,fpgu}. This  allows FPGAs to offer comparable performance to GPUs while retaining the advantage of power efficiency and architectural flexibility. Additionally, by leveraging coarse-grained architectures and dedicated Digital Signal Processing (DSP) slices, FPGAs can efficiently handle data parallelism and computationally intensive workloads, making them increasingly suitable for ML applications \cite{gemmini}, \cite{architecture_symposium}.

An exciting recent development in accelerating training times is speculative backpropagation \cite{gim}, \cite{simultaneous}. This technique seeks to overlap the forward pass and backward pass of neural network training, using speculative weight updates when error gradients fall within a certain threshold. By storing past error gradients and reusing them when subsequent gradients are within a defined tolerance, speculative backpropagation reduces the need for repetitive computations, thereby significantly decreasing training time without substantially compromising accuracy.

In this work, we implement the MNIST dataset \cite{mnist} using OpenMP as our parallel programming platform. OpenMP allows us to take advantage of speculative backpropagation by concurrent execution of forward and backward passes. By utilizing OpenMP's multi-threading capabilities, we significantly reduce training time while maintaining accuracy. As a next step, the application will be synthesized on a state-of-the-art FPGA. The current implementation and evaluation have been carried out on CPUs to assess the performance benefits of speculative backpropagation prior to full hardware deployment. Additionally, we provide a visualization for speculative backpropagation to better understand its impact on performance and accuracy.

In summary, the key contributions of this work include:-
\begin{itemize}
  \item Demonstrating the performance gains achievable through speculative backpropagation on FPGA-based accelerators.
  \item Analyzing the trade-offs between execution time reduction and accuracy degradation for varying speculative thresholds. 
\end{itemize}

The rest of this paper is organized as follows: Section \ref{sec:back} provides a brief background of neural networks and speculative backpropagation approach. Section \ref{sec:method} provides our implementation details and presents the experimental results. Section \ref{sec:challenges} discusses potential challenges and future directions. Finally Section \ref{sec:Conclusion} concludes this paper. We have open-sourced all our work here\footnote{https://github.com/Gitsc592/Speculative\_Backprop\_Project}.

\section{Background}
\label{sec:back}


The Internet of Things (IoT) market has been growing phenomenally over the past decade and is poised for significant growth in the coming years. As of 2023 there were approximately 17 billion IoT connected devices globally and this is expected to grow to 19 billion within a year at a 13\% rate \cite{iot}. The IoT market size is expected to reach approximately $\$4,000$ billion by 2032 with a CAGR of approximately 24\% \cite{fortune}. Cellular IoT connections were projected to reach 4 billion by the end of last year \cite{IoTcon}. Additionally the Industrial IoT market is expected to grow to approximately $\$390$ billion by 2028 at a CAGR of 22\% \cite{bcc}.
Amongst the top barriers for IoT growth are cybersecurity concerns, power consumption and battery life issues. At the same time, ML applications are rapidly expanding within the IoT field. This creates a need to optimize neural networks, which play a central role in these applications. Several studies have investigated different approaches to achieve these optimizations \cite{book:gdrChapter8, patent:simar,tesla_array}.

\begin{figure} [h]
    \centering
    \includegraphics[width=1\linewidth]{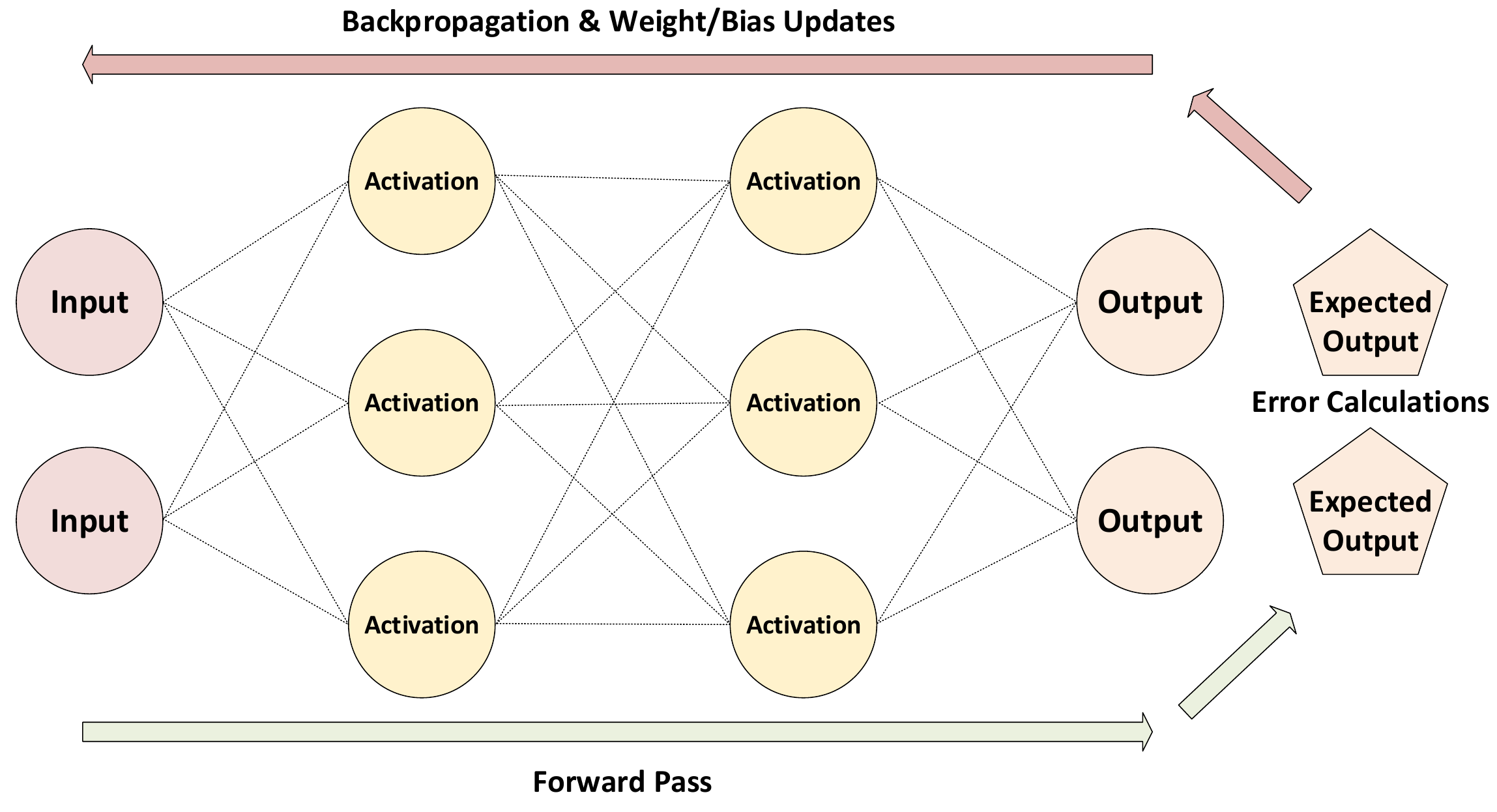}
    \caption{A fully connected neural network architecture illustrating forward pass, error calculation, and backpropagation}
    \label{fig:nn}
\end{figure}

\subsection{Neural Nets}


A neural network is a ML algorithm composed of layers of interconnected nodes, where each connection is defined by a weight.  Each node has an associated bias and an activation value. These weights and biases represent the connection between nodes and how they influence the network. The network begins with an input layer that receives raw data and ends with an output layer, which produces the final predictions. Fig. \ref{fig:nn} illustrates a fully connected network neural network architecture. 


There are three sequences of operations that a neural network goes through when training the weights and bias values: \textit{ forward pass, backward pass}, and \textit{weight update}. The forward pass takes a set of inputs and passes it through the network. The activations and weights are multiplied and summed to calculate the activation on the next layer's corresponding node. The bias of the node is also added to the summation. An activation function is applied to introduce nonlinearity \cite{blog:gsReproducibility}. This cycle continues until the output layer returns a set of activations.

During the backward pass, the previous output from a forward pass is used to compute an error value. This error value is used to recursively compute error gradients for each weight and bias term \cite{Widrow30}. The final phase being the weight updates, applies the error gradients to the weights and bias, adjusting them for future computations \cite{book:gdrChapter8}. 

\subsection{Current approach}
\begin{figure}
    \centering
    \includegraphics[width=1\linewidth]{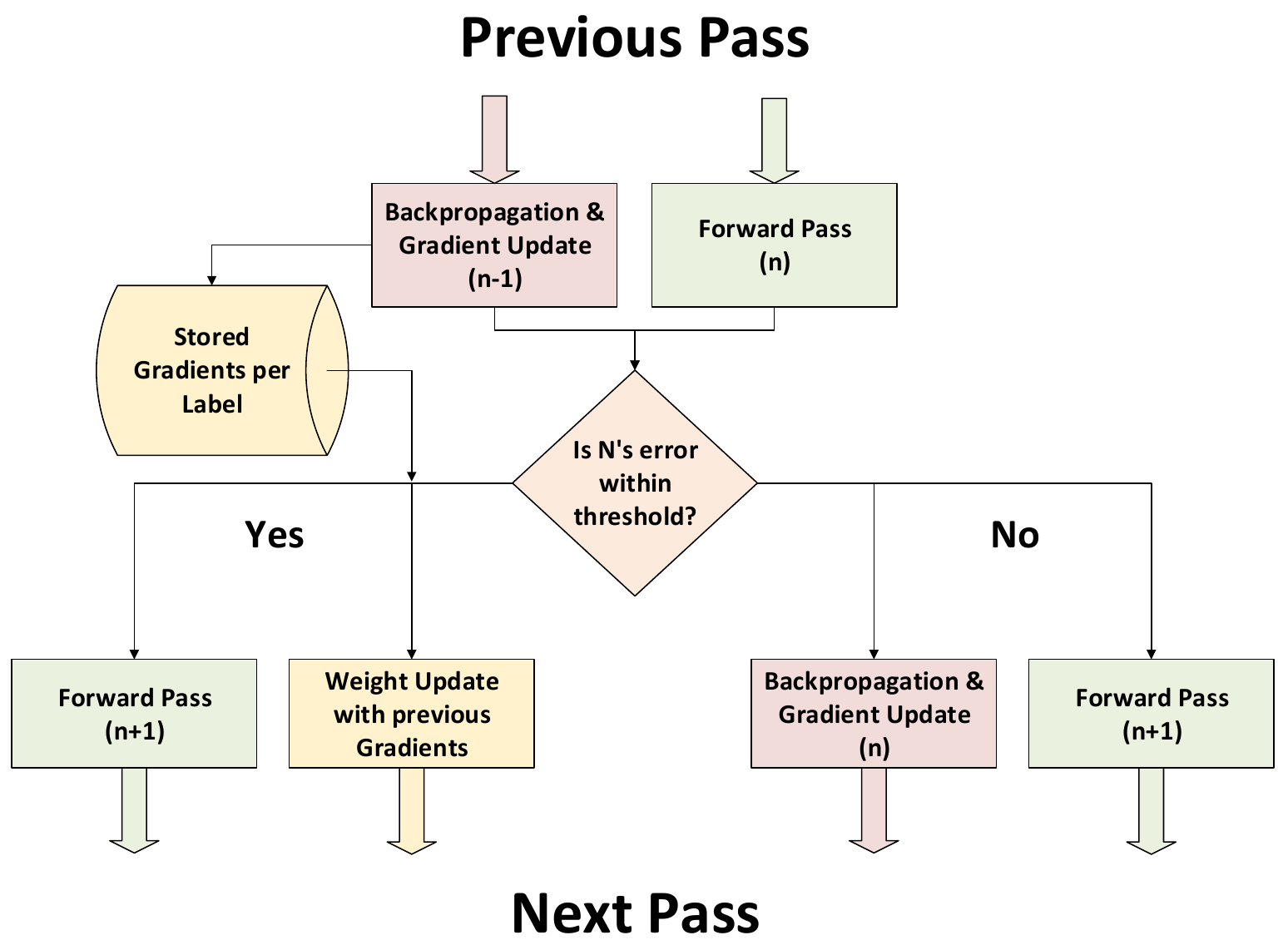}
    \caption{Speculative backpropagation: reuse gradients within a threshold to overlap forward and backward passes}
    \label{fig:spec}
\end{figure}

Traditional neural network carry out the process of forward pass and backpropagation  sequentially. Parallelizing these steps could significantly enhance the algorithm’s performance. However, since each phase relies on outputs from the previous layer, applying parallelism directly is challenging.
Speculative backpropagation overcomes this limitation by using values from the previous forward pass during the backpropagation phase. This allows both the forward pass and backpropagation to execute simultaneously, since they no longer depend on the same set of values. However, the weight update phase must still occur after both are complete \cite{simultaneous}. Fig. \ref{fig:spec} gives a visual representation of how speculative backpropagation works.

\subsection{Speculative backpropagation}

The core idea behind speculative backpropagation is to reuse previously computed error gradients when the difference between output layers fall within a specified threshold. By storing past gradients for specific labels, the algorithm can bypass recalculating the backward pass if the output similarity meets this criterion. This allows saving computation time. If the outputs fall outside the threshold, a standard backpropagation step is triggered, and the gradients are updated accordingly \cite{simultaneous}.



\section{Methodology}
\label{sec:method}




Combining the speculative backpropagation approach with task parallelism (thread-level) has the potential to significantly accelerate neural network training. In our implementation, the OpenMP programming platform was used to enable simultaneous execution of the forward and backward passes by running them on separate threads. Synchronization occurs once both tasks are completed, ensuring proper progression of the training process.

\subsection{Architecture}

To evaluate the effectiveness of speculative backpropagation, we implemented a simple feedforward neural network and trained it on the MNIST dataset \cite{mnist}. The architecture consists of 4 layers: an input layer, 2 hidden layers, and an output layer. Each hidden layer contains 16 neurons that use the ReLU activation function. The output layer uses Softmax\cite{simultaneous}. These choices help preserve useful gradient flow during training.\cite{simultaneous}. Table \ref{neural} shows the configuration of the neural network.



\begin{table}[h]
\centering
\begin{tabular}{|c|c|c|}
\hline
\textbf{Layer}   & \textbf{Neurons} & \textbf{Activation Function} \\ \hline
Input            & 784              & Pixel Value / 255                  \\ \hline
Hidden Layer \#1 & 16               & Leaky ReLU                   \\ \hline
Hidden Layer \#2 & 16               & Leaky ReLU                   \\ \hline
Output           & 10               & Softmax                      \\ \hline
\end{tabular}
\caption{Neural Network Configuration}
\label{neural}
\end{table}

The ReLU activation function used is the leaky variant, which scales inputs below zero by 0.01 to prevent neurons from becoming inactive and unable to contribute to weight updates. To avoid gradient explosions during both standard and speculative backpropagation, gradients are clipped between -5 and 5. A learning rate of 0.01 is applied, meaning gradients are scaled before updating weights to prevent overshooting near local minima. The batch size is set to 15, allowing the network to accumulate gradients over 15 samples before performing a weight update.



\subsection{Tools}

The neural network was implemented in the C programming language and synthesized using the Xilinx Vitis IDE and Xilinx Vivado. Initial development and testing was performed on an Intel i7-8665U CPU clocked at 2.4 GHz. The target hardware platform for deployment is the Octavo OSDZU3-REF FPGA-based accelerator \cite{web:octavo_specs}, which is a work-in-progress.

\begin{figure*}[t]
    \centering
    \begin{subfigure}[t]{0.33\textwidth}
        \centering
        \includegraphics[width=\linewidth]{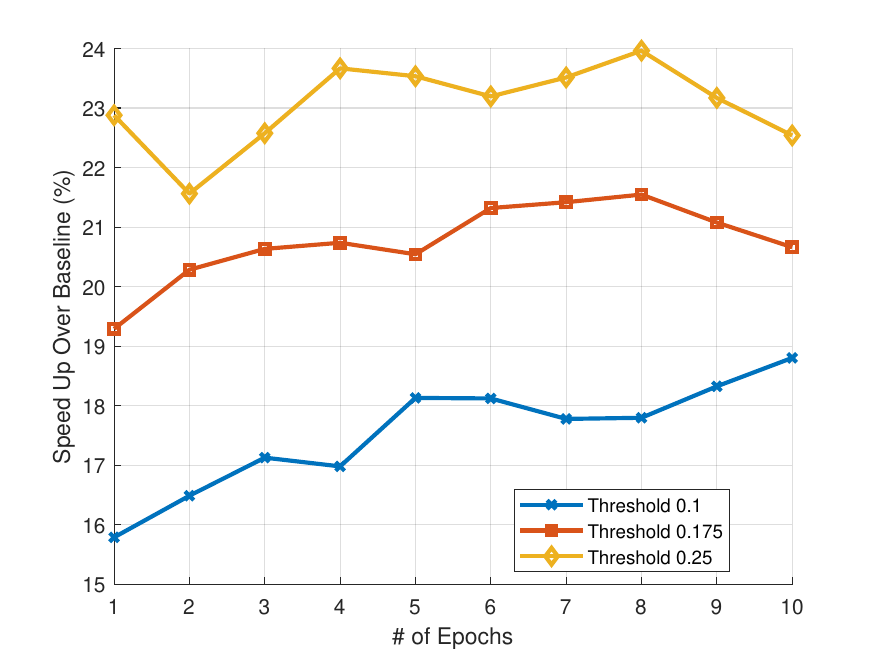}
        \caption{Execution time speedup over baseline}
        \label{fig:execution}
    \end{subfigure}%
    \begin{subfigure}[t]{0.33\textwidth}
        \centering
        \includegraphics[width=\linewidth]{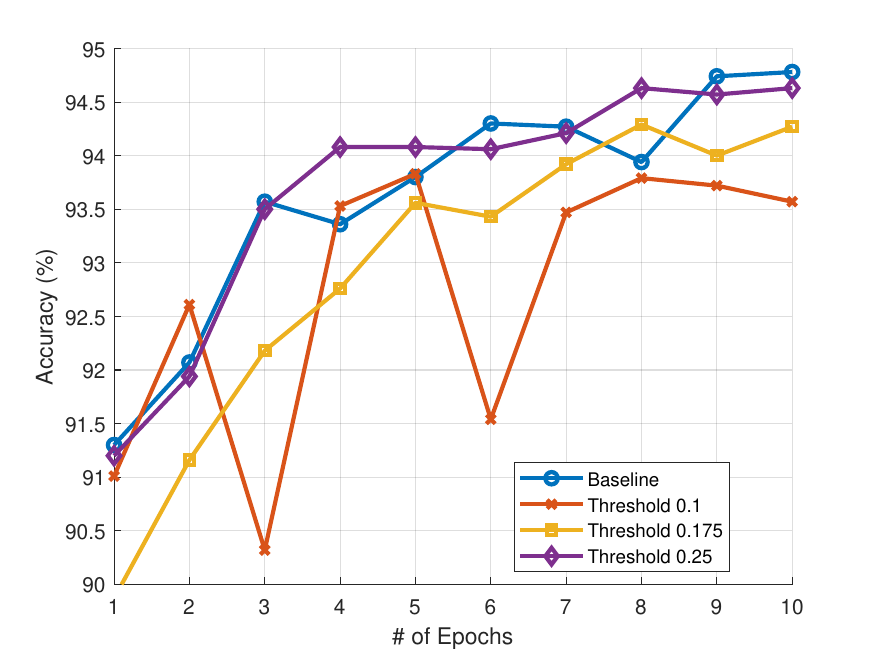}
        \caption{Epochs vs Accuracy}
        \label{fig:accuracy}
    \end{subfigure}%
    \begin{subfigure}[t]{0.33\textwidth}
        \centering
        \includegraphics[width=\linewidth]{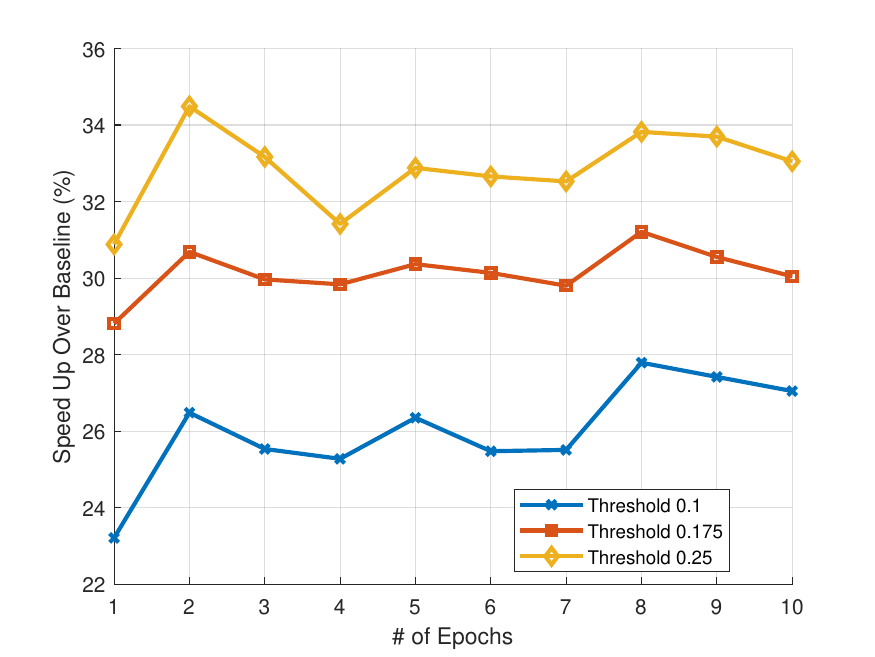}
        \caption{Propagation step execution time speedup over baseline}
        \label{fig:step graph}
    \end{subfigure}
    \caption{Results}
    \label{fig:Results}
\end{figure*}



\begin{table}[]
\centering
\begin{tabular}{|ccccc|}
\hline
\multicolumn{5}{|c|}{\textbf{Execution   Time (sec)}}                                                                                                                 \\ \hline
\multicolumn{1}{|c|}{\textbf{Epochs}} & \multicolumn{1}{c|}{\textbf{Baseline}} & \multicolumn{1}{c|}{\textbf{Th. 0.1}} & \multicolumn{1}{c|}{\textbf{Th. 0.175}} & \textbf{Th. 0.25} \\ \hline
\multicolumn{1}{|c|}{1}               & \multicolumn{1}{c|}{11.44}             & \multicolumn{1}{c|}{9.88}         & \multicolumn{1}{c|}{9.59}           & 9.31          \\ \hline
\multicolumn{1}{|c|}{2}               & \multicolumn{1}{c|}{22.89}             & \multicolumn{1}{c|}{19.65}        & \multicolumn{1}{c|}{19.03}          & 18.83         \\ \hline
\multicolumn{1}{|c|}{3}               & \multicolumn{1}{c|}{34.26}             & \multicolumn{1}{c|}{29.25}        & \multicolumn{1}{c|}{28.40}          & 27.95         \\ \hline
\multicolumn{1}{|c|}{4}               & \multicolumn{1}{c|}{45.88}             & \multicolumn{1}{c|}{39.22}        & \multicolumn{1}{c|}{38.00}          & 37.10         \\ \hline
\multicolumn{1}{|c|}{5}               & \multicolumn{1}{c|}{57.27}             & \multicolumn{1}{c|}{48.48}        & \multicolumn{1}{c|}{47.51}          & 46.36         \\ \hline
\multicolumn{1}{|c|}{6}               & \multicolumn{1}{c|}{68.57}             & \multicolumn{1}{c|}{58.05}        & \multicolumn{1}{c|}{56.52}          & 55.66         \\ \hline
\multicolumn{1}{|c|}{7}               & \multicolumn{1}{c|}{80.16}             & \multicolumn{1}{c|}{68.06}        & \multicolumn{1}{c|}{66.02}          & 64.90         \\ \hline
\multicolumn{1}{|c|}{8}               & \multicolumn{1}{c|}{91.67}             & \multicolumn{1}{c|}{77.82}        & \multicolumn{1}{c|}{75.42}          & 73.95         \\ \hline
\multicolumn{1}{|c|}{9}               & \multicolumn{1}{c|}{102.60}            & \multicolumn{1}{c|}{86.71}        & \multicolumn{1}{c|}{84.74}          & 83.30         \\ \hline
\multicolumn{1}{|c|}{10}              & \multicolumn{1}{c|}{114.17}            & \multicolumn{1}{c|}{96.10}        & \multicolumn{1}{c|}{94.62}          & 93.17         \\ \hline
\end{tabular}
\caption{Raw Data for Training Execution Time of Speculative Backpropagation over Baseline}
\label{table:Exe Time}
\end{table}

\begin{table}[]
\centering
\begin{tabular}{|ccccc|}
\hline
\multicolumn{5}{|c|}{\textbf{Accuracy   (\%)}}\\ \hline
\multicolumn{1}{|c|}{\textbf{Epochs}} & \multicolumn{1}{c|}{\textbf{Baseline}} & \multicolumn{1}{c|}{\textbf{Th. 0.1}} & \multicolumn{1}{c|}{\textbf{Th. 0.175}} & \textbf{Th. 0.25} \\ \hline
\multicolumn{1}{|c|}{1}               & \multicolumn{1}{c|}{91.30\%}           & \multicolumn{1}{c|}{91.01\%}                & \multicolumn{1}{c|}{89.87\%}                  & 91.20\%                 \\ \hline
\multicolumn{1}{|c|}{2}               & \multicolumn{1}{c|}{92.07\%}           & \multicolumn{1}{c|}{92.61\%}                & \multicolumn{1}{c|}{91.16\%}                  & 91.94\%                 \\ \hline
\multicolumn{1}{|c|}{3}               & \multicolumn{1}{c|}{93.57\%}           & \multicolumn{1}{c|}{90.32\%}                & \multicolumn{1}{c|}{92.18\%}                  & 93.50\%                 \\ \hline
\multicolumn{1}{|c|}{4}               & \multicolumn{1}{c|}{93.36\%}           & \multicolumn{1}{c|}{93.53\%}                & \multicolumn{1}{c|}{92.76\%}                  & 94.08\%                 \\ \hline
\multicolumn{1}{|c|}{5}               & \multicolumn{1}{c|}{93.80\%}           & \multicolumn{1}{c|}{93.83\%}                & \multicolumn{1}{c|}{93.56\%}                  & 94.08\%                 \\ \hline
\multicolumn{1}{|c|}{6}               & \multicolumn{1}{c|}{94.30\%}           & \multicolumn{1}{c|}{91.54\%}                & \multicolumn{1}{c|}{93.43\%}                  & 94.06\%                 \\ \hline
\multicolumn{1}{|c|}{7}               & \multicolumn{1}{c|}{94.27\%}           & \multicolumn{1}{c|}{93.47\%}                & \multicolumn{1}{c|}{93.92\%}                  & 94.21\%                 \\ \hline
\multicolumn{1}{|c|}{8}               & \multicolumn{1}{c|}{93.94\%}           & \multicolumn{1}{c|}{93.79\%}                & \multicolumn{1}{c|}{94.29\%}                  & 94.63\%                 \\ \hline
\multicolumn{1}{|c|}{9}               & \multicolumn{1}{c|}{94.74\%}           & \multicolumn{1}{c|}{93.72\%}                & \multicolumn{1}{c|}{94.00\%}                  & 94.57\%                 \\ \hline
\multicolumn{1}{|c|}{10}              & \multicolumn{1}{c|}{94.78\%}           & \multicolumn{1}{c|}{93.57\%}                & \multicolumn{1}{c|}{94.27\%}                  & 94.63\%                 \\ \hline
\end{tabular}
\caption{Raw Data for Accuracy of Speculative Backpropagation and Baseline}
\label{table:Accuracy}
\end{table}

\begin{table}[]
\centering
\begin{tabular}{|ccccc|}
\hline
\multicolumn{5}{|c|}{\textbf{Execution   Time for Individual Step (usec)}}                                                                                                           \\ \hline
\multicolumn{1}{|c|}{\textbf{Epochs}} & \multicolumn{1}{c|}{\textbf{Baseline}} & \multicolumn{1}{c|}{\textbf{Th. 0.1}} & \multicolumn{1}{c|}{\textbf{Th. 0.175}} & \textbf{Th. 0.25} \\ \hline
\multicolumn{1}{|c|}{1}               & \multicolumn{1}{c|}{181.69}            & \multicolumn{1}{c|}{147.46}           & \multicolumn{1}{c|}{141.05}             & 138.82            \\ \hline
\multicolumn{1}{|c|}{2}               & \multicolumn{1}{c|}{184.49}            & \multicolumn{1}{c|}{145.86}           & \multicolumn{1}{c|}{141.17}             & 137.18            \\ \hline
\multicolumn{1}{|c|}{3}               & \multicolumn{1}{c|}{182.33}            & \multicolumn{1}{c|}{145.24}           & \multicolumn{1}{c|}{140.29}             & 136.92            \\ \hline
\multicolumn{1}{|c|}{4}               & \multicolumn{1}{c|}{181.63}            & \multicolumn{1}{c|}{144.98}           & \multicolumn{1}{c|}{139.89}             & 138.21            \\ \hline
\multicolumn{1}{|c|}{5}               & \multicolumn{1}{c|}{181.82}            & \multicolumn{1}{c|}{143.90}           & \multicolumn{1}{c|}{139.47}             & 136.83            \\ \hline
\multicolumn{1}{|c|}{6}               & \multicolumn{1}{c|}{181.58}            & \multicolumn{1}{c|}{144.71}           & \multicolumn{1}{c|}{139.53}             & 136.88            \\ \hline
\multicolumn{1}{|c|}{7}               & \multicolumn{1}{c|}{181.19}            & \multicolumn{1}{c|}{144.36}           & \multicolumn{1}{c|}{139.59}             & 136.72            \\ \hline
\multicolumn{1}{|c|}{8}               & \multicolumn{1}{c|}{196.34}            & \multicolumn{1}{c|}{143.14}           & \multicolumn{1}{c|}{139.41}             & 136.69            \\ \hline
\multicolumn{1}{|c|}{9}               & \multicolumn{1}{c|}{186.46}            & \multicolumn{1}{c|}{142.93}           & \multicolumn{1}{c|}{139.50}             & 136.22            \\ \hline
\multicolumn{1}{|c|}{10}              & \multicolumn{1}{c|}{181.44}            & \multicolumn{1}{c|}{142.81}           & \multicolumn{1}{c|}{139.52}             & 136.37            \\ \hline
\end{tabular}
\caption{Raw Data for Average Execution Time for each Forward and Speculative Backward Propagation step over Baseline}
\label{table:step table}
\end{table}

\subsection{Results} 
This section presents the performance improvements achieved through speculative backpropagation with parallelism approach. 

Fig. \ref{fig:execution} shows the speed up of the neural network with speculative backpropagation compared to baseline implementation. Each point on the x-axis represents the number of epochs completed. For instance, an epoch step of 1 indicates the program ran for a single epoch, while step 5 corresponds to five completed epochs. The overall speed up is the result of performing the forward and backward propagation steps in parallel. A slight upward trend in speedup is observed as the number of epochs increases, attributed to the growing similarity between previous and current forward-propagated configurations \cite{simultaneous}. This increases the likelihood of speculative backpropagation being triggered, as more outputs fall within the predefined threshold. 

Fig. \ref{fig:accuracy} compares the accuracy between the different configurations of speculative backpropagation. The graph shows minimal accuracy loss, never straying more than \textit{3-4\%} away from baseline. This also results in previous forward propagated configurations becoming similar to current ones due to smaller differences between configurations\cite{simultaneous}. Tables \ref{table:Exe Time} and \ref{table:Accuracy} present the exact raw values for training execution time and accuracy, respectively, comparing different threshold configurations of speculative backpropagation against the baseline.

    




Finally, Fig. \ref{fig:step graph} shows the speed up in average execution time of a speculative propagation step over the baseline. The time is taken over 1 step of the forward and backward propagation task excluding gradient application and storing previous values and averaged over all 60000 training images multiplied by the amount of epochs. The graph shows a maximum of \textit{35\%} speed up in execution time over the baseline with a threshold of 0.25. This shows the same increasing trend for the same reasons discussed. The trend is more apparent in Table \ref{table:step table}, which shows the raw data with exact values.

\section{Challenges and Future Directions}
\label{sec:challenges}

While speculative backpropagation offers promising performance gains, its effectiveness depends on the similarity between successive forward pass outputs. This may limit its applicability in tasks involving highly variable inputs, such as broad image classification across unrelated object categories (e.g., cars vs. animals). Additionally, speculative updates require storing past gradients, which increases memory usage. This can become a bottleneck in large-scale or edge-deployed models with limited storage.

Our current work uses a simple MLP architecture trained on the MNIST dataset to evaluate speculative backpropagation. We plan to extend this to convolutional neural networks (CNNs) to assess its effectiveness in deeper, spatially structured models. We also aim to complete synthesis and deployment on the Octavo OSDZU3-REF FPGA platform. Mapping the approach to RTL will require dedicated forward and speculative backward modules, along with control logic for threshold comparison.

Finally, future work may explore dynamic thresholding strategies and selective gradient retention to reduce memory overhead while maintaining speedups. These enhancements aim to make speculative backpropagation more robust, scalable, and hardware-efficient—especially in real-time and edge AI applications where power and latency are critical.  Additional work includes improvements on memory localization and deeper pipelining \cite{gim}, \cite{gimIPDPSW2024}, and \cite{report:ggdr}.

\section{Conclusion}
\label{sec:Conclusion}

This work demonstrates the potential of speculative backpropagation to accelerate neural network training by overlapping the forward and backward passes. Utilizing OpenMP as the parallel programming platform enables simultaneous execution of these phases, significantly reducing training time. Experimental results indicate that speculative backpropagation with a threshold of 0.25 achieves a maximum speedup of 24\% in execution time compared to the baseline, while maintaining accuracy within 3-4\%. Additionally, a 35\% speedup over baseline was observed for individual propagation steps, highlighting the effectiveness of overlapping computations. Future work can explore applying this approach to more complex datasets and larger, deeper neural network architectures. 

\bibliographystyle{IEEEtran}
\bibliography{ieee}

\begin{thebibliography}{10}
\providecommand{\url}[1]{#1}
\csname url@samestyle\endcsname
\providecommand{\newblock}{\relax}
\providecommand{\bibinfo}[2]{#2}
\providecommand{\BIBentrySTDinterwordspacing}{\spaceskip=0pt\relax}
\providecommand{\BIBentryALTinterwordstretchfactor}{4}
\providecommand{\BIBentryALTinterwordspacing}{\spaceskip=\fontdimen2\font plus
\BIBentryALTinterwordstretchfactor\fontdimen3\font minus
  \fontdimen4\font\relax}
\providecommand{\BIBforeignlanguage}[2]{{%
\expandafter\ifx\csname l@#1\endcsname\relax
\typeout{** WARNING: IEEEtran.bst: No hyphenation pattern has been}%
\typeout{** loaded for the language `#1'. Using the pattern for}%
\typeout{** the default language instead.}%
\else
\language=\csname l@#1\endcsname
\fi
#2}}
\providecommand{\BIBdecl}{\relax}
\BIBdecl

\bibitem{socctaxonomy}
A.~A. Purkayastha, S.~A. Shiddhibhavi, and H.~Tabkhi, ``Taxonomy of spatial
  parallelism on fpgas for massively parallel applications,'' in \emph{2018
  31st IEEE International System-on-Chip Conference (SOCC)}, 2018, pp. 55--60.

\bibitem{dnn}
\BIBentryALTinterwordspacing
A.~Nechi, L.~Groth, S.~Mulhem, F.~Merchant, R.~Buchty, and M.~Berekovic,
  ``Fpga-based deep learning inference accelerators: Where are we standing?''
  \emph{ACM Trans. Reconfigurable Technol. Syst.}, vol.~16, no.~4, Oct. 2023.
  [Online]. Available: \url{https://doi.org/10.1145/3613963}
\BIBentrySTDinterwordspacing

\bibitem{fpgu}
\BIBentryALTinterwordspacing
M.~Al~Kadi, B.~Janssen, and M.~Huebner, ``Fgpu: An simt-architecture for
  fpgas,'' in \emph{Proceedings of the 2016 ACM/SIGDA International Symposium
  on Field-Programmable Gate Arrays}, ser. FPGA '16.\hskip 1em plus 0.5em minus
  0.4em\relax New York, NY, USA: Association for Computing Machinery, 2016, p.
  254–263. [Online]. Available: \url{https://doi.org/10.1145/2847263.2847273}
\BIBentrySTDinterwordspacing

\bibitem{gemmini}
H.~Genc, ``Gemmini: An agile systolic array generator enabling systematic
  evaluations of deep-learning architectures,'' \emph{DAC}, 2021.

\bibitem{architecture_symposium}
Jouppi, ``In-datacenter performance analysis of a tensor processing unit,''
  \emph{Proceedings of the 44th Annual International Symposium on Computer
  Architecture, ISCA '17'}, pp. 1--12, 2017.

\bibitem{gim}
M.~Borowicz, J.~Ding, W.~Fan, Z.~Gao, D.~Jackson, A.~Lu, S.~Rohlfsen, and
  R.~Simar, ``Gim (ghost in the machine): A dsp-inspired accelerator platform
  for exploring machine-learning architectures,'' in \emph{2024 IEEE 67th
  International Midwest Symposium on Circuits and Systems (MWSCAS)}, 2024, pp.
  1120--1124.

\bibitem{simultaneous}
S.~Park, ``Speculative backpropagation for cnn parallel training,'' 2020.

\bibitem{mnist}
``Mnist dataset - gtdlbench,'' accessed: 2025-03-29.

\bibitem{iot}
``Number of connected iot devices growing 13\% to 18.8 billion,''
  \url{https://iot-analytics.com/number-connected-iot-devices/}, accessed:
  2025-03-30.

\bibitem{fortune}
``Internet of things [iot] market size, share, growth, trends, 2032,''
  \url{https://www.fortunebusinessinsights.com/industry-reports/internet-of-things-iot-market-100307},
  accessed: 2025-03-30.

\bibitem{IoTcon}
``Iot connections forecast ericsson mobility report,''
  \url{https://www.ericsson.com/en/reports-and-papers/mobility-report/dataforecasts/iot-connections-outlook},
  accessed: 2025-03-30.

\bibitem{bcc}
M.~Kudle, ``Global industrial iot market size and growth forecast report,''
  \url{https://www.bccresearch.com/market-research/information-technology/industrial-iot-market.html},
  accessed: 2025-03-30.

\bibitem{book:gdrChapter8}
R.~W. D.E.~Rumelhart, G.E.~Hinton, ``Chapter 8: Learning internal
  representations by error propagation,'' 1986.

\bibitem{patent:simar}
R.~Simar, ``Learning device and method,'' \emph{US Patent 5,109,351}, 1992.

\bibitem{tesla_array}
K.~A.~H. P.~J.~Bannon and E.~Talpes, ``Accelerated mathematical engine,''
  \emph{US Patent App. 15/710}, Jan. 24 2019.

\bibitem{blog:gsReproducibility}
\BIBentryALTinterwordspacing
D.~L. G.~Shamir, ``Reproducibility in deep learning and smooth activations,''
  \emph{Google Research Blog}, 2023. [Online]. Available:
  \url{https://blog.research.google/2022/04/reproducibility-in-deep-learning-and.html}
\BIBentrySTDinterwordspacing

\bibitem{Widrow30}
B.~Widrow and M.~Lehr, ``30 years of adaptive neural networks: Perceptron,
  madaline, and backpropagation,'' \emph{Proceedings of the IEEE}, vol.~78,
  no.~9, pp. 1415--1442, 1990.

\bibitem{web:octavo_specs}
\BIBentryALTinterwordspacing
``Octavo system osdzu3 system-in-package,'' 2024. [Online]. Available:
  \url{https://octavosystems.com/octavo\_products/osdzu3/}
\BIBentrySTDinterwordspacing

\bibitem{gimIPDPSW2024}
\BIBentryALTinterwordspacing
M.~Borowicz, J.~Ding, W.~Fan, Z.~Gao, D.~Jackson, A.~Lu, S.~Rohlfsen, and
  R.~Simar, ``Gim (ghost in the machine): A coarse-grained reconfigurable
  compute-in-memory platform for exploring machine-learning architectures,''
  \emph{2024 IEEE International Parallel and Distributed Processing Symposium
  Workshops}, 2024. [Online]. Available:
  \url{https://doi.org/10.1109/IPDPSW63119.2024.00126}
\BIBentrySTDinterwordspacing

\bibitem{report:ggdr}
R.~Simar, ``Generalizing the generalized delta-rule: Architectures and
  alternatives,'' \emph{Texas Instruments Internal Technical Report}, 1987.

\end{thebibliography}

\end{document}